\documentclass{article}

\usepackage{PRIMEarxiv}

\usepackage[utf8]{inputenc}
\usepackage[T1]{fontenc}
\usepackage{hyperref}

\RequirePackage{doi}

\usepackage{url}
\usepackage{xurl}

\usepackage{amsmath,amssymb,amsfonts}
\usepackage{graphicx}
\graphicspath{{media/}} 
\usepackage{booktabs}
\usepackage{array}
\usepackage{tabularx}
\usepackage{longtable}
\usepackage{pdflscape}
\usepackage{adjustbox}
\usepackage{subcaption}

\usepackage{xcolor}
\usepackage{tikz}
\usepackage{pgfplots}
\pgfplotsset{compat=1.18}
\usetikzlibrary{shapes.geometric, arrows, arrows.meta, positioning}

\usepackage{listings}
\usepackage{todonotes}
\usepackage[final]{microtype}
\usepackage{nicefrac}

\usepackage[authoryear,round]{natbib}

\makeatletter
\@ifpackageloaded{hyperref}{}{%
  \RequirePackage[colorlinks=true, linkcolor=black, citecolor=black, urlcolor=black]{hyperref}}
\@ifpackageloaded{xurl}{}{\RequirePackage{xurl}}
\@ifpackageloaded{biblatex}{%
  \ExecuteBibliographyOptions{doi=true}%
  \DeclareFieldFormat{doi}{%
    \href{https://doi.org/\detokenize{#1}}{doi:\space #1}}%
}{}

\providecommand{\doi}[1]{%
  \href{https://doi.org/\detokenize{#1}}{\nolinkurl{#1}}}
\makeatother

\usepackage{fancyhdr}
\definecolor{primary}{HTML}{1F77B4}
\definecolor{secondary}{HTML}{AEC7E8}
\definecolor{tertiary}{HTML}{FF7F0E}
\definecolor{accent}{HTML}{2CA02C}
\definecolor{textcolor}{HTML}{333333}

\definecolor{codegray}{rgb}{0.5,0.5,0.5}
\definecolor{codepurple}{rgb}{0.58,0,0.82}
\definecolor{backcolour}{rgb}{0.95,0.95,0.92}

\lstdefinestyle{mystyle}{
    backgroundcolor=\color{backcolour},
    commentstyle=\color{codegray},
    keywordstyle=\color{blue},
    numberstyle=\tiny\color{codegray},
    stringstyle=\color{codepurple},
    basicstyle=\ttfamily\footnotesize,
    breakatwhitespace=false,
    breaklines=true,
    captionpos=b,
    keepspaces=true,
    numbers=left,
    numbersep=5pt,
    showspaces=false,
    showstringspaces=false,
    showtabs=false,
    tabsize=2
}
\tikzstyle{startstop} = [rectangle, rounded corners, minimum width=3cm, minimum height=1cm,text centered, draw=black, fill=red!30]
\tikzstyle{process}   = [rectangle, minimum width=3cm, minimum height=1cm, text centered, draw=black, fill=blue!30]
\tikzstyle{decision}  = [diamond, minimum width=3cm, minimum height=1cm, text centered, draw=black, fill=green!30]
\tikzstyle{arrow}     = [thick,->,>=stealth]

\title{Digital Twins for Intelligent Intersections:\\ A Literature Review}

\author{
  Alben Rome Bagabaldo, Jürgen Hackl \\
  Department of Civil and Environmental Engineering \\
  Princeton University \\
  Princeton, NJ, USA 08544\\
  \texttt{\{alben, hackl\}@princeton.edu} \\
}

\begin{document}
\maketitle

\begin{abstract}
Intelligent intersections play a pivotal role in urban mobility, demanding innovative solutions such as digital twins to enhance safety and efficiency. This literature review investigates the integration and application of digital twins for intelligent intersections, a critical area within smart urban traffic systems. The review systematically categorizes existing research into five key thematic areas: (i) Digital Twin Architectures and Frameworks; (ii) Data Processing and Simulation Techniques; (iii) Artificial Intelligence and Machine Learning for Adaptive Traffic Control; (iv) Safety and Protection of Vulnerable Road Users; and (v) Scaling from Localized Intersections to Citywide Traffic Networks. Each theme is explored comprehensively, highlighting significant advancements, current challenges, and critical insights. The findings reveal that multi-layered digital twin architectures incorporating real-time data fusion and AI-driven decision-making enhances traffic efficiency and safety. Advanced simulation techniques combined with sophisticated AI/ML algorithms demonstrate notable improvements in real-time responsiveness and predictive accuracy for traffic management. Additionally, the integration of digital twins has shown substantial promise in safeguarding vulnerable road users through proactive and adaptive safety strategies. Despite these advancements, key challenges persist, including interoperability of diverse data sources, scalability of digital twins for extensive traffic networks, and managing uncertainty within dynamic urban environments. Addressing these challenges will be essential for the future development and deployment of intelligent, adaptive, and sustainable intersection management systems.
\end{abstract}

\keywords{digital twins \and traffic management \and artificial intelligence \and safety \and smart infrastructure \and intersections \and review}

\section{Introduction}
\label{sec:intro}

Road infrastructure plays a foundational role in modern society by connecting communities, driving economic growth, and supporting access to essential services. In the United States alone, public roads span approximately 4.19 million miles \citep{FHWA2023HM18} and include over 15.8 million intersections \citep{Streetsblog2022Deadliest}, with 255 million drivers undertaking around 227 billion driving trips daily and spending an average of 60.2 minutes on the road \citep{steinbach2023american}. Despite their ubiquity, intersections account for a disproportionate share of traffic incidents, responsible for over 40\% of crashes \citep{Azimi2014STIP, Namazi2019Intelligent, Choi2010CrashFactors} and 27\% of traffic fatalities \citep{NHTSA2024Traffic}. High rates of intersection-related crashes stem from factors such as driver errors, poor visibility, and adverse weather conditions, while pedestrians face an even greater risk, with pedestrian fatalities rising by 75\% since 2010 \citep{SGA2024Dangerous}. In 2023, fatal and severe traffic crashes in the U.S. resulted in \$1.85 trillion in total losses, including \$460 billion in direct economic costs and \$1.4 trillion in quality-of-life impacts \citep{TRIP2024Addressing}.

There are many attempts to address these underlying issues for traffic intersections, including modernization of signal systems and geometric improvements \citep{Chandler2013Signalized}. Among these, digital twin (DT) technology has emerged as particularly promising \citep{Tao2019Digital}. A digital twin represents a virtual counterpart of a physical system or process, continuously updated through real-time sensor data, simulations, and algorithms~\citep{di2024ai}. The concept of digital twins was first introduced in 2003 within Michael Grieves’s Product Lifecycle Management framework at University of Michigan \citep{Grieves2014DigitalTwin}. Subsequent adoption of the term by NASA in ``Digital Twin Paradigm for Future NASA and U.S. Air Force Vehicles'' \citep{Glaessgen2012Digitala} further spurred interest in this area. Since then, digital twins have provided a powerful framework to address challenges, such as reducing crash risk, improving pedestrian safety, mitigating driver errors, and adapting to adverse weather conditions, by leveraging real-time sensor data, advanced simulation tools, and machine learning algorithms.

Virtual models that mirror physical intersections allow stakeholders to predict traffic behavior, identify hazards, and implement adaptive traffic control strategies \citep{Abdelrahman2025What, NHTSA2024Traffic}. Moreover, this innovation has the potential to significantly improve safety for vulnerable road users (VRUs), such as pedestrians, cyclists, and micromobility device users~\citep{Townsend2023Summary}. By offering a dynamic, data-driven approach to analyze and optimize intersection performance, digital twins can transform urban traffic systems to be safer, more efficient, and more sustainable~\citep{xu2025enhancing}.

\section{Methodology}
\label{sec:methodology}

This section outlines our structured, multi-stage review process, modeled after best practices in systematic literature reviews, to assemble and analyze research on digital twins for intelligent intersections. We organized papers into five themes mentioned in Section~\ref{sec:intro}-Introduction.

Parallel to the developments in digital twins, the significant growth of scholarly literature has driven the development of innovative approaches to synthesizing research findings. Recent studies have shown that although academics are reading more articles than ever before, the average time spent per article has decreased, indicating a shift toward more superficial engagement with the literature~\citep{Tenopir2015Scholarly}. To manage this increasing information load, LLMs have emerged as valuable tools to automate and expedite systematic literature reviews. For instance, LLMs can efficiently perform tasks such as title and abstract screening and preliminary data synthesis, thereby reducing the human workload and enhancing the comprehensiveness of reviews~\citep{Dennstadt2024Title, Luo2024Potential}. Despite concerns regarding potential hallucinations and the need for human oversight~\citep{Chelli2024Hallucination, Khraisha2024Can}, the integration of LLMs represents a significant methodological advancement. In our work, we leverage these capabilities to systematically integrate insights on digital twin applications in intelligent intersections.

\subsection{Literature Search and Screening}

We queried ScienceDirect using broad terms: \textit{``digital twins,'' ``intelligent intersections,'' and ``transportation''} for publications from January 2015 to May 2025 and selecting only papers that are Open Access. Following database retrieval, all full texts were processed for automated summarization using Gemini 2.0 Pro, which ingests complete PDFs and outputs structured summaries under a schema capturing scope, methods, intersection relevance, and thematic signals.

Each summary was screened by asking the LLM prompt: \textit{“Is this paper relevant to digital twins for traffic intersections?”} with rationale. Outputs were then manually reviewed. Papers without relevance to cross roads or intersections or focused outside transportation were excluded. To enhance thematic depth, we manually inspected references from all included papers. Works presenting foundational frameworks or significant empirical advances were incorporated through a snowballing process, which extended the corpus beyond database searches to include both seminal and emergent contributions. This snowballing process includes those papers that were identified from the supplementary thematic search described in subsection~\ref{sec:theme-search}.

\begin{figure}[ht]
\centering
\begin{tikzpicture}
\begin{axis}[
    ybar,
    bar width=20pt,
    ymin=0,
    ymax=40,
    ylabel={Number of Papers},
    xlabel={Publication Year},
    xtick=data,
    nodes near coords,
    nodes near coords align={vertical},
    symbolic x coords={2019,2020,2021,2022,2023,2024,2025},
    xticklabel style={font=\small, rotate=45, anchor=east},
    axis lines=left,
    tick style={line width=0.5pt},
    tick align=outside,
    every axis plot/.style={fill=blue!30, draw=blue},
    major tick length=4pt,
    enlarge x limits={abs=0.8cm}  
]
\addplot coordinates {
    (2019,1)
    (2020,5)
    (2021,2)
    (2022,8)
    (2023,26)
    (2024,34)
    (2025,11)
};
\end{axis}
\end{tikzpicture}
\caption{Number of papers by publication year.}
\label{fig:no-of-papers}
\end{figure}
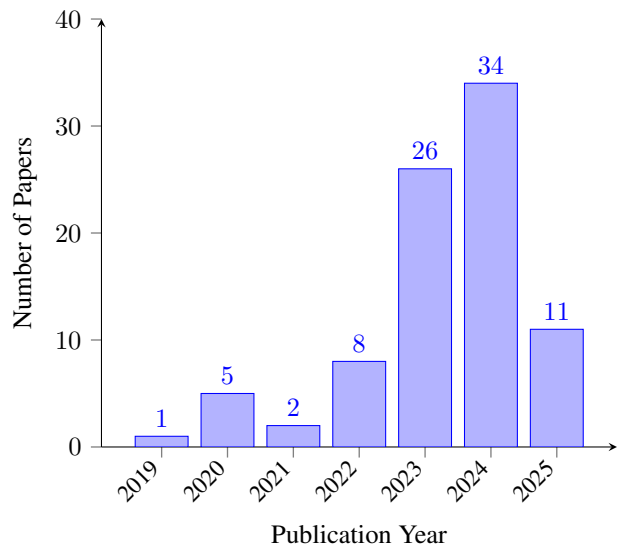

During the search, we initially identified 544 papers, this number includes the papers identified in the snowballing process. After applying LLM-based filtering and skimming through the content, this number was reduced to 85 unique articles, which were included in this review. As shown in Figure~\ref{fig:no-of-papers}, the distribution of papers by publication year reveals a clear upward trend. From 2019 to 2022, between 1 and 6 relevant studies were identified annually (1 in 2019, 5 in 2020, 2 in 2021, and 8 in 2022). A remarkable increase begins in 2023 with 26 papers, followed by a further rise to 34 papers in 2024. Although the literature search was cut off in May 2025, 11 papers were already identified for that year, indicating that 2025 is on pace for continued substantial contributions. This trend highlights the increasing significance and rapid development of this research domain.

\subsection{Thematic Classification and Keyword-Augmented Searches}
\label{sec:theme-search}

All these relevant papers were manually classified into the aforementioned themes. We assigned each paper to the theme reflecting its primary contribution. Many papers overlapped themes, and were documented accordingly. To ensure comprehensive coverage, we also conducted theme-specific supplementary searches using refined keyword strings or keywords shown in Table~\ref{tab:themes}.

\begin{table}[ht]
  \centering
  \small
  \caption{Themes, additional keywords used, and number of unique papers considered.}
  \label{tab:themes}
  \begin{tabularx}{\textwidth}{|l|X|c|}
    \hline
    \textbf{Theme} & \textbf{Keywords} & \textbf{Final No. of Papers} \\
    \hline
    Digital Twin Architectures and Frameworks &
      digital twins architecture, interoperability, digital twin design & 18 \\
    \hline
    Data Processing and Simulation Techniques &
      data fusion, sensor integration, simulation & 19 \\
    \hline
    AI/ML in Traffic Control &
      reinforcement learning traffic, AI signal control, multi-agent control, deep learning & 23 \\
    \hline
    Safety and Protection of Vulnerable Road Users &
      pedestrian digital twin, vulnerable road user safety, agent-based safety simulation, ethics in traffic AI & 15 \\
    \hline
    Scaling to Citywide Traffic Networks &
      urban scale digital twins, distributed control, smart-city infrastructure, IoT-enabled twins & 35 \\
    \hline
  \end{tabularx}
\end{table}

This process led to the inclusion of additional papers with distinct relevance to each theme. We organized the papers into theme-specific folders and used Gemini 2.0 Pro to generate initial syntheses for each category. Several studies contributed to more than one theme, reflecting the interdisciplinary and interconnected nature of digital twin research in transportation. The number of papers associated with each theme is as follows: 18, 19, 23, 15, and 35, respectively as shown in Table~\ref{tab:themes}. The synthesis prompts guided the extraction of shared findings, key technical advancements, current limitations, and emerging research directions. These LLM-generated summaries served as thematic scaffolding, which we subsequently refined through manual review, writing, and editing into the coherent narrative sections presented in this paper.

\section{Digital Twin Architectures and Frameworks} 
\label{sec:dt-architectures}

The concepts of architectures and frameworks are important in structuring and implementing effective digital twins. A digital twin architecture defines the structured organization of components and their relationships within a digital twin system. This architecture is essential for enabling interoperability and data exchange with both the physical system and external applications~\citep{Cavalieri2023Proposal}. Frameworks, on the other hand, constitute the conceptual and architectural foundation that supports the integration of heterogeneous data sources, computational infrastructures, and analytical models. They offer standardized practices that enhance interoperability and scalability across a range of applications~\citep{Nativi2021Digital}.

To address the question, \textit{``What architectural and interoperability frameworks are necessary to develop scalable, modular, and interoperable digital twins for intelligent intersections?''}, we conducted a thorough search and filtering process and identified 18 scholarly articles specifically focused on digital twin architectures and frameworks relevant to the topic. These selected studies offer valuable insights into current practices and innovations in the field. The search parameters and inclusion criteria are detailed in Section \ref{sec:methodology} – Methodology.

This section surveys the architectural foundations and enabling technologies driving digital twins for intelligent intersections. The discussion is organized into two subsections: \textit{Architectures}, which covers real-time synchronization mechanisms, data-driven and optimization-centric systems, AI-enhanced structures, and domain-specific configurations; and \textit{Frameworks and Enabling Technologies}, which explores supporting computational methods, interoperability protocols, and deployment strategies that facilitate scalable and adaptive digital twin implementations. Together, these approaches illustrate the diverse, multi-layered, and rapidly evolving ecosystem of digital twin solutions for traffic management and intersection intelligence.

\subsection{Architectures}

Recent advancements in digital twin architectures showcase diverse system designs that integrate real-time sensing, communication protocols, and simulation tools to support intelligent intersections. A zero calibration, satellite ground mapping architecture is presented in literature, which fuses hierarchical spatiotemporal video analysis, on-the-fly 3D bounding box estimation, and deep convolutional neural networks (CNNs) to support real-time environment modeling~\citep{Rezaei20233DNeta}. Likewise, a continuously synchronized framework is adopted, combining live traffic feeds, vehicle-to-everything (V2X) protocols such as Dedicated Short-Range Communications (DSRC)~\citep{Kenney2011Dedicateda}, Wireless Access in Vehicular Environments (WAVE)~\citep{Jiang2019Wireless}, and Cellular V2X (C-V2X) over 5G New Radio (the fifth generation cellular standard), along with a Simulation of Urban MObility (SUMO)-based microscopic simulator~\citep{Lopez2018Microscopic} via the Traffic Control Interface (TraCI) API~\citep{Wegener2008TraCI} \citep{Kusic2023digitala}.

Hardware-in-the-Loop (HIL) simulation and mixed reality virtual layers further refine control logic~\citep{Wagner2023SPaTa}, while another work integrates roadside units (RSUs), millimeter-wave (mmWave) radar, Light Detection and Ranging (LiDAR), weather stations, and edge computing microservices into a cloud native co-simulation platform~\citep{An2023Roadsidea}. Co-simulation refers to the coordinated execution of two or more simulation tools, often running in parallel and exchanging data in real time, to realistically represent complex, multi-domain systems such as traffic flow and vehicle behavior~\citep{Yilmaz-Niewerth2024Developinga}.

An intelligent digital twin system is outlined, featuring integrated sensor data, a Remote Dictionary Server (Redis) database backbone~\citep{Redis2025}, and Kalman-filtered prediction modules. This architecture enables client-side visualization while avoiding reliance on direct simulation engines, supporting a lightweight forecasting layer~\citep{Liu2024Developmenta}. Redis is a high-performance, open-source, in-memory NoSQL data structure store that can be used as a database, cache, and message broker.

Data-driven and optimization-centric architecture emphasize model accuracy, robustness to missing data, and automated decision-making. A digital twin is presented that leverages Gaussian mixture clustering, expectation–maximization (EM) algorithms, regression analysis, and singular value decomposition (SVD) for iterative engine-performance indexing~\citep{Taghavi2024Advanced}. Meanwhile, a temporal neighboring interpolation and missing data imputation routines to reconstruct sparse traffic volumes in connected corridors is also introduced \citep{Saroj2023Impacta}. In the rail domain, a method is presented that automatically generates track geometry from airborne LiDAR point clouds using Industry Foundation Classes (IFC)-based parametric assemblies and mesh-based reconstruction. The dataset also includes challenging scenarios such as where a railway line crosses a road, demonstrating the system’s ability to handle complex geometries \citep{Ariyachandra2023Leveraging}. Meanwhile, an architecture is structured around a monitoring and visualization dashboard that feeds an Optimal Level Crossing (OLC) optimization model, supported by a physical-to-virtual data layer that ensures safe rail crossing sequences \citep{Djordjevic2024optimisationbaseda}.

Certain literature also focuses on layered AI and generative frameworks, coupling foundational models with feedback loops. A three-layer digital twin is synthesized, comprising an AI core, bidirectional physical-to-virtual and virtual-to-physical channels for sensor ingestion and actuator commands, and RL agents for adaptive control~\citep{Kreuzer2024Artificiala}. Building on this, a GenAI-enabled large-flow model is introduced that unifies Generative Adversarial Networks (GANs), Variational Autoencoders (VAEs), transformers, diffusion models, and foundation models for adaptive urban flow simulation~\citep{Huang2025Generativea}. Another approach integrates graph neural networks (GNNs) and spatial–temporal graph attention networks (STGATs) to blend static infrastructure features with dynamic traffic states, driving preventive maintenance assessments rather than continuous clustering or optimization routines~\citep{Lu2025Modeling}. On the other hand, a pipeline style architectural framework is proposed, centered on context-aware segmentation of LIDAR point clouds, 3D meshing, semantic labeling, and asset management integration, prioritizing geometric fidelity over deep generative reasoning \citep{Davletshina2024Automatinga}.

Domain specific communication and representation frameworks further diversify digital twin architectures. A dual vehicle digital twin system (head and assistant) is implemented using V2X reservation intervals and Operational Design Domain (ODD) constraints \citep{Du2024Digital}. Another implementation embeds SPaT/MAP protocols within a PLC-based control and HIL simulation environment, augmented by mixed-reality visualization \citep{Wagner2023SPaTa}. A complementary cloud-native digital twin architecture integrates edge computing, multi-modal sensing, and open service APIs to manage distributed sensor arrays in support of vehicle-infrastructure cooperative intelligent driving \citep{An2023Roadsidea}. Finally, a scan-to-graph pipeline converts segmented point clouds through lane polynomial fitting and Frenet transformations into a graph representation, offering a graph-centric perspective on highway digital twins \citep{Pan2024Scantographa}.

\begin{figure}[h]
  \centering
  \includegraphics[width=\textwidth]{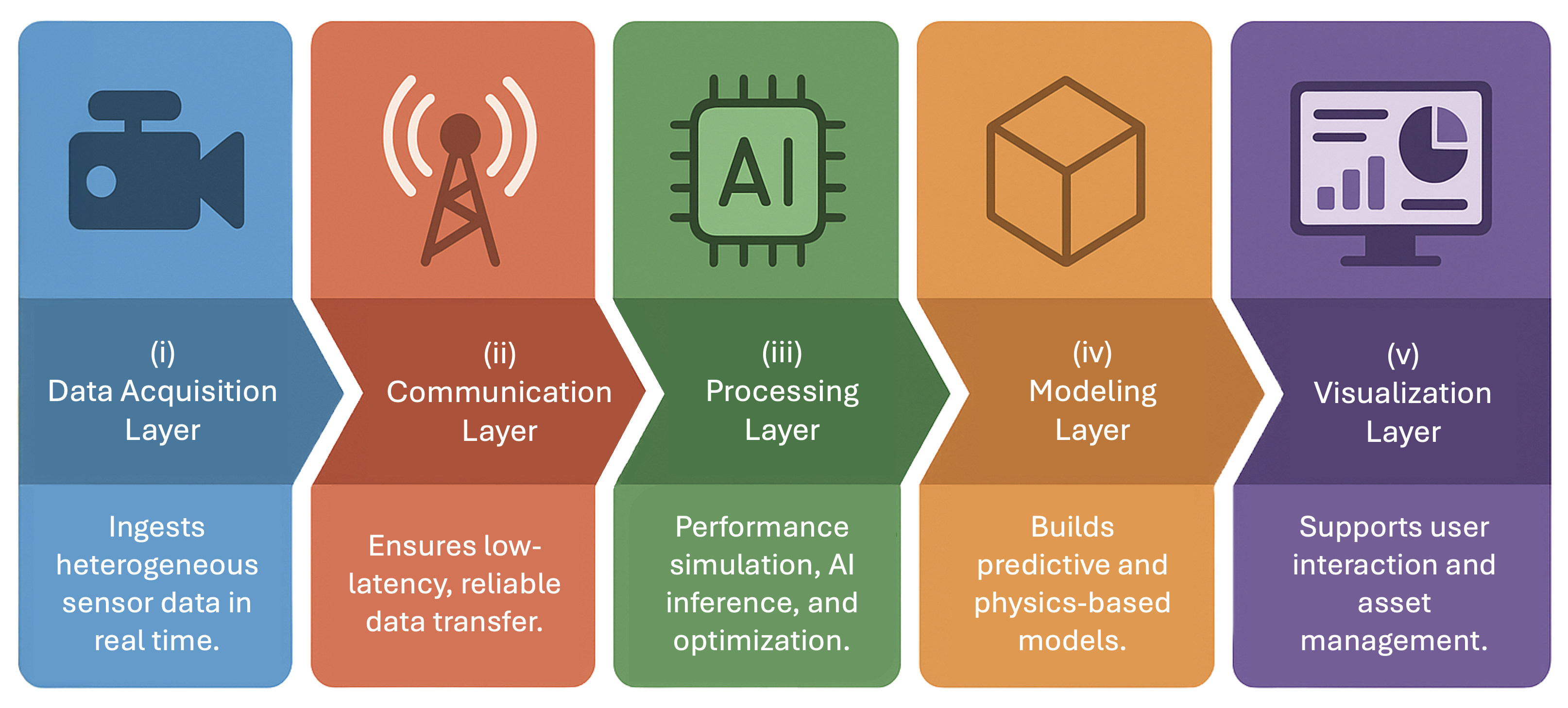}
  \caption{Layered digital twin architecture for intelligent intersections, shown left to right: 
  \textit{(i) Data Acquisition Layer} ingests heterogeneous real-time inputs (e.g., video, LiDAR, radar, loop detectors, weather stations); 
  \textit{(ii) Communication Layer} ensures low-latency, reliable transfer via technologies such as DSRC/WAVE, C-V2X, 5G NR, MQTT, and edge microservices; 
  \textit{(iii) Processing Layer} performs simulation, AI inference, data fusion and optimization (e.g., SUMO, GMM \& EM, GNNs, generative models); 
  \textit{(iv) Modeling Layer} constructs predictive and physics-informed representations including foundation models, mesh reconstruction, and parametric assemblies; 
  and \textit{(v) Visualization Layer} supports operator interaction and asset management through web dashboards, BIM viewers, and mixed-reality interfaces. 
  The directional flow highlights how raw sensing progresses through communication and computation into actionable models and user-facing insights.}
  \label{fig:dt_multilayer_architecture}
\end{figure}

Many surveyed studies adopt a multilayer architecture, where \textit{layer} refers to a distinct functional component within the system's design, each responsible for specific tasks. The architecture can be divided into five key layers: (i) the data acquisition layer, which ingests heterogeneous sensor streams such as video, LiDAR, radar, loop detectors, and weather stations \citep{Rezaei20233DNeta, Kusic2023digitala, Pan2024Scantographa}; (ii) the communication layer, which leverages V2X protocols (DSRC/WAVE, C-V2X), 5G New Radio, Message Queuing Telemetry Transport (MQTT), and edge microservices for low-latency data transfer \citep{Wagner2023SPaTa, Du2024Digital, Aguiar2022MobiWise}; (iii) the processing layer, which employs simulation engines such as SUMO \citep{Kusic2023digitala} and HIL setups \citep{Wagner2023SPaTa}, clustering and optimization algorithms including Gaussian Mixture Models(GMMs) and EM \citep{Taghavi2024Advanced} and Optimal Level Crossing optimization \citep{Djordjevic2024optimisationbaseda}, AI modules such as GNNs and STGATs \citep{Lu2025Modeling}, GANs and VAEs \citep{Huang2025Generativea}, and RL agents for adaptive control \citep{Kreuzer2024Artificiala}; (iv) the modeling layer, which integrates physics-based solvers, foundation-model reasoning, and parametric assemblies like IFC and mesh-based reconstruction to support predictive analytics and maintenance planning \citep{Liu2024Developmenta}; and (v) the visualization layer, which delivers web dashboards, Building Information Modeling (BIM) interfaces, and mixed-reality environments for user interaction and asset management \citep{Sabato2023BIM, An2023Roadsidea, Wagner2023SPaTa}.
This multilayered architecture is shown in the Figure \ref{fig:dt_multilayer_architecture}.

\subsection{Frameworks and Enabling Technologies}

The diverse architectures presented in the literature are enabled by frameworks that coordinate interactions between layers, components, and technologies. These frameworks typically incorporate specific computational, communication, and coordination technologies.

A key enabler across these frameworks is the integration of AI/ML techniques. These are often employed not only for data analysis and imputation, such as GMMs, expectation–maximization, and GNNs, but also for real-time decision-making and predictive control. For instance, clustering and regression techniques are used to generate robust engine performance indices in maritime digital twins through GMMs, EM algorithms, and singular value decomposition~\citep{Taghavi2024Advanced}. In the urban traffic domain, generative AI models unify GANs, VAEs, and transformers to simulate adaptive traffic flow under varying city conditions, offering a scalable and sustainable approach to mobility forecasting~\citep{Huang2025Generativea}. AI-enhanced digital twin architectures for smart intersections also employ RL and context-aware control mechanisms to coordinate infrastructure and vehicle behavior in real time~\citep{Wagner2023SPaTa}. In support of road asset maintenance, STGATs and GNNs are integrated to fuse infrastructure features and traffic states for condition prediction and preventive maintenance~\citep{Lu2025Modeling}. Finally, deep learning and context-aware segmentation pipelines are applied to point cloud data to automate the geometric modeling of road assets, prioritizing meshing accuracy and semantic labeling over generative simulation~\citep{Davletshina2024Automatinga}.

Edge computing and cloud-native deployment strategies are also central to many frameworks. These enable distributed processing of sensor data and the scalability needed for citywide deployments. Technologies such as Docker containers, MQTT protocols (a lightweight, publish-subscribe network protocol designed for connecting remote devices with resource constraints or limited network bandwidth), and microservice-based architectures support modularity, fault tolerance, and efficient data handling. For instance, one system utilizes edge computing microservices and multi-modal sensing to manage distributed data \citep{An2023Roadsidea} , while another one integrates a Redis database and Kalman filter-based forecasting to enable lightweight, real-time traffic predictions \citep{Liu2024Developmenta}. Similarly, the MobiWise architecture provides a microservice-based, IoT-leveraged framework for eco-routing decisions using MQTT and containerized services \citep{Aguiar2022MobiWise}. Another foundational element is the use of standardized data formats and interoperability protocols. Tools like SUMO and the TraCI API are used to ensure consistent simulation and control integration, as demonstrated in real-time traffic synchronization frameworks \citep{Kusic2023digitala}.

Visualization frameworks play a pivotal role in enabling interaction with digital twins. These range from web dashboards to immersive mixed-reality interfaces and BIM-integrated viewers. Systems such as BIM-integrated visualization for smart intersections offer intuitive interfaces for real-time monitoring, coordination, and decision-making \citep{Sabato2023BIM}.

Beyond conventional architectural and control-oriented perspectives, recent work has also introduced the concept of ``Fused Twins,'' which refers to digital twins that are embedded within their physical environments through situated analytics. These twins integrate the physical, data, analytical, virtual, and connection environments into in-situ overlays, enabling real-world objects to be augmented with digital layers for enhanced context-awareness. This approach encourages embodied interfaces, such as AR/VR overlays that are spatially grounded in the physical world, and calls for empirical studies evaluating cognitive benefits such as reduced mental workload in real deployments \citep{Grubel2022Hitchhikers}. In transportation, a related survey of digital twin applications for autonomous driving proposed a test framework based on V2X communication for real-time simulation, but underscored the lack of clear definitions and real-world validation in current digital twin-based testbeds \citep{Niaz2021Autonomousa}.

\section{Data Processing and Simulation Techniques}
\label{sec:dt-data-sim}

Effective digital twins for intelligent intersections rely on two interconnected pillars: (i) robust data processing pipelines and (ii) high fidelity simulation platforms. Data processing pipelines must integrate and fuse heterogeneous sensor streams, such as loop detectors, LiDAR and radar scanners, and connected vehicle broadcasts, through data ingestion, cleaning, and fusion steps. Standardizing and combining these diverse inputs in real time enables big data analytics to forecast traffic conditions and support traffic demand management~\citep{Torre-Bastida2018Big}. Simulation techniques span microscopic traffic models (e.g., SUMO~\citep{Lopez2018Microscopic}), dynamic traffic assignment~\citep{Wang2018Dynamic}, and co-simulation frameworks that integrate autonomous vehicle (AV) simulators (e.g., CARLA~\citep{Dosovitskiy17CARLA}) with traffic flow models~\citep{Varga2023Optimizinga,Yilmaz-Niewerth2024Developinga}. These platforms also support calibration and validation of virtual sensor models and AI-driven synthetic data generation for enhanced digital twin realism.

This section synthesizes 19 key studies that explore data processing and simulation strategies essential for developing digital twins for intelligent intersections, with a focus on addressing the question: \textit{``What data fusion strategies and uncertainty modeling approaches can enhance the reliability, accuracy, and real-time decision-making capabilities of digital twins?''} Table~\ref{tab:dt-data-sim-techniques} (in the appendices) contrasts chosen digital twin works based on their data fusion or processing approaches and corresponding simulation environments.

The remainder of this section is organized into two subsections. \textit{Real-Time Data Fusion and Integration} focuses on heterogeneous data fusion methods, including techniques for combining sensor inputs such as loop detectors, LiDAR, radar, and connected vehicle broadcasts. It also highlights structured fusion frameworks, missing data restoration, and predictive modules that support proactive traffic management. \textit{Simulation Platforms} examines digital twin environments, emphasizing co-simulation, calibration and validation of virtual sensor models, and AI-driven synthetic data generation.

\subsection{Real-time Data Integration and Heterogeneous Data Fusion}
\label{sec:datafusion}

Digital twins for intelligent intersections rely on a diverse range of data sources to faithfully replicate real-world dynamics and support decision-making in real time. These sources include structured data from loop detectors, LiDAR and radar sensors, GPS traces from connected vehicles, traffic signal states, and simulation logs from platforms such as SUMO and CARLA~\citep{Kusic2023digitala,Yilmaz-Niewerth2024Developinga,Aguiar2022MobiWise}. Unstructured and semi-structured data, such as textual incident reports, GIS metadata, and social movement patterns, are increasingly fused with sensor inputs to capture contextual and behavioral dimensions of urban mobility~\citep{Katsumbe2024Extractiona,Villanueva-Merino2024Leveraging}. Additionally, high fidelity perception data, such as annotated images, point clouds, and acoustic signals, enhance simulation realism and feed machine learning models for predictive analytics~\citep{Lv2022Memoryaugmenteda,Bamminger2023Evaluationa,Yadav2025Leasta}. Examples of data collection and ingestion include the low-latency integration of heterogeneous sources such as loop detectors, radar, and floating car data into unified twin environments~\citep{Kusic2023digitala}, while real-time energy analytics and eco-routing rely on dynamic script-driven updates within simulation platforms like SUMO~\citep{Angelina2024Introducing,Aguiar2022MobiWise}. These data streams typically undergo preprocessing steps such as cleaning, normalization, and transformation to enable downstream tasks like forecasting and decision support. Despite the richness of these inputs, missing or incomplete data remains a persistent challenge. Sensor outages, transmission delays, or coverage gaps can degrade the performance of digital twins by introducing uncertainty into traffic flow predictions and safety assessments. To address this, recent studies have applied techniques such as k-means clustering and temporal-neighbor interpolation to restore missing traffic volume counts with measurable improvements in output fidelity~\citep{Saroj2023Impacta}. Structured fusion frameworks and forecasting models also play a critical role in mitigating data sparsity by inferring plausible values and enabling proactive system responses~\citep{Louati2025Machinea,Hassani2024systematic}.

As outlined in the preceding section, a foundational pillar of digital twins for intelligent intersections is the integration and fusion of heterogeneous data streams in real time. Systems continuously ingest roadside sensor and vehicle data into unified, low-latency twin environments~\citep{Kusic2023digitala,Aguiar2022MobiWise}. Data gaps in traffic volume counts can be characterized and imputed (e.g., through $k$-means clustering of loss patterns and temporal-neighbor interpolation) to maintain application performance~\citep{Saroj2023Impacta}. Energy analytics integrated with SUMO via Python scripts enable estimation of consumption under varying scenarios~\citep{Angelina2024Introducing}. Pavement management twins merge heterogeneous logs using spatiotemporal graph attention networks for condition prediction~\citep{Lu2025Modeling}. Event detection can be performed in a lightweight, online manner by combining PCA with DBSCAN to identify accidents from GPS traces~\citep{Papadopoulos2024Lightweighta}. In parallel, Logistics 4.0 applications employ connected tracking and tracing devices to provide real-time, end-to-end visibility~\citep{Helo2024Logistics}.

Structured fusion frameworks further support decision-making. Common strategies, like complementary, redundant, and cooperative fusion, operate at the signal/raw, feature/intermediate, and decision levels, often organized under established taxonomies such as the Joint Directors of Laboratories (JDL), Luo–Kay, or Dasarathy models~\citep{White1991Data, Hassani2024systematic}. Beyond sensor streams, unstructured text can be mined via natural language processing (NLP) methods (e.g., Word2Vec) to extract salient system elements and relationships for urban and transport modeling, subsequently informing quantitative indicators~\citep{Katsumbe2024Extractiona}. Visual fusion in transportation twins leverages memory-augmented modules to segment evolving scenes, particularly in self-driving contexts~\citep{Lv2022Memoryaugmenteda}. Cross-domain pipelines unify IoT inputs, multi-objective optimization outputs, and microservice architectures, often coupled to simulation twins, to address data sparsity~\citep{Aguiar2022MobiWise}. Hybrid approaches pair statistical tools with machine learning methods (e.g., PCA with DBSCAN for online event detection and ARIMA/ANN for forecasting safety and energy trends) to synthesize diverse indicators into coherent insights~\citep{Papadopoulos2024Lightweighta, Louati2025Machinea}.

\subsection{Simulation Platforms}
\label{sec:simulation}

Intersection twins often require concurrent traffic and AV simulations. Digital twins of intersections increasingly depend on co-simulation frameworks that synchronize traffic flow models with AV simulators to test control, sensing, and coordination strategies. For example, CARLA and SUMO are linked in lock step co-simulation by exchanging vehicle state vectors at each time step to analyze AV performance under mixed traffic conditions \citep{Yilmaz-Niewerth2024Developinga}. To scale corridor-level traffic studies, mesoscopic dynamic traffic assignment (DTA) engines run in tandem with microscopic SUMO routines, allowing efficient signal-timing evaluation across large urban networks \citep{Kuraksin2020Integrateda}. For energy-aware modeling, detailed consumption profiles are scripted into SUMO using fine-grained variables \citep{Angelina2024Introducing}. Specialized testbeds like CupCarbon support ambient noise stress testing, employing least squares–predicted acoustic signals and multi-criteria asynchronous classifiers to simulate smart accident management \citep{Yadav2025Leasta}. Digital twins for eco-routing leverage real-time IoT loops within the MobiWise framework, feeding sensor data into microscopic simulators for continual optimization \citep{Aguiar2022MobiWise}. To ensure perceptual realism, high fidelity sensor models (LiDAR, radar, camera) are validated in multi-body simulation platforms based on experimental benchmarks \citep{Bamminger2023Evaluationa}. Meanwhile, simulation performance tuning is explored through downscaling strategies that introduce mesoscopic traffic simulation to offload non-critical traffic dynamics from microscopic AV simulators, improving computation time by up to 500\%~\citep{Varga2023Optimizinga}. Communication standards are also integrated: a co-simulation testbed coupling SUMO, Veins, and OMNeT++ validates SPaT/MAP V2X protocols between traffic lights and vehicles to support intersection control in connected environments \citep{Wagner2023SPaTa}. Even path-following control laws are tested through co-simulation environments like CarSim and MATLAB/Simulink; for instance, the Pure Pursuit algorithm has been optimized using proportional-derivative (PD) error corrections to reduce overshoot on high-curvature segments \citep{Cao2025Optimization}.

High fidelity perception within twins requires rigorous calibration and validation. Building on the need for accurate simulation of intersection dynamics, advanced perception modeling ensures that digital twins can faithfully represent real-world conditions. High-resolution perception models are calibrated using field-recorded data and simulation platforms such as CarMaker, where tools like Scenario Record,Replay,Rearrange (RRR) enable injection of real scenarios; this is combined with Mixture Density Networks to close the sim-to-reality gap by over 30\%\citep{Bamminger2023Evaluationa}. Memory-augmented segmentation networks allow digital twins to self-tune visual fidelity by aligning outputs with ground-truth image sequences, dynamically adapting to environmental changes\citep{Lv2022Memoryaugmenteda}. Pavement twins further exploit STGATs to harmonize predicted road states with sensor observations in highway infrastructure management~\citep{Lu2025Modeling}. Urban-scale pedestrian flow models are refined using social simulation kernels that blend GIS datasets with movement patterns of local populations, enabling inclusive digital twin testing for age-friendly infrastructure~\citep{Villanueva-Merino2024Leveraging}. In acoustics, asynchronous sound recordings from field deployments are used to validate smart accident detection models driven by multi-criteria classifiers embedded within CupCarbon simulators~\citep{Yadav2025Leasta}.

\section{Artificial Intelligence and Machine Learning in Traffic Control}
\label{sec:ai-ml}

Leveraging digital twin architectures and continuous data assimilation discussed in Sections ~\ref{sec:dt-architectures} and \ref{sec:dt-data-sim}, we can see how AI/ML further empower these twins to improve traffic control.
By integrating AI/ML algorithms, these digital replicas, which serve as virtual mirrors of physical systems, are empowered to simulate and intelligently analyze and predict evolving traffic conditions.

In the context of intelligent intersections, a digital twin uses live data from vehicles, traffic signals, and sensors to dynamically map the real world into a virtual environment, ingesting multiple sources such as vehicle counts, speeds, signal states, and V2X messages~\citep{Lv2022Memoryaugmented, Mi2024benchmark}.  
This integration enables enhanced visualization, adaptive decision-making, and optimized performance in real time, allowing operators or autonomous controllers to react swiftly to fluctuations in demand.
Within the virtual model, scenarios can be tested and optimized continuously; for example, signal timing strategies or lane configurations can be tested before deployment, and the digital twin can predict queue buildups or spillovers to enable preventive adjustments to signals or traveler information systems.

Traditional traffic control relies on fixed-time or rule-based plans that cannot adjust to real-time fluctuations. In contrast, AI-powered adaptive systems digest streaming data from multiple sources and make split-second decisions to optimize flow, reduce congestion, and improve safety~\citep{Aulia2024new}.  
Techniques such as RL and neural network–based prediction enable controllers to learn optimal responses (e.g., when to extend a green light or trigger an alert) from experience and data rather than static programs, addressing the complexity of urban traffic where patterns change hourly and incidents require immediate response~\citep{Lai2024Reinforcement, Yao2023Advanced}.

This section addresses the question, \textit{``How can AI/ML methods, such as RL and DL, enhance adaptive traffic control within digital twins?''}
This synthesis covers 23 recent studies and is structured into four subsections: \textit{Traditional ML Methods for Traffic Control}, which discusses early AI applications such as random forests, SVMs, and time series models for forecasting and classification; \textit{RL and Multi-Agent Coordination}, which explores adaptive signal control using RL agents and collaborative strategies for network-level optimization; \textit{Deep Learning for Prediction, Perception, and Control}, which highlights CNNs, GNNs, and hybrid models for spatio-temporal forecasting and incident detection; and \textit{AI/ML-Driven Edge-to-Cloud Architectures and Immersive Decision Interfaces}, which presents emerging paradigms including distributed inference, quantum-accelerated optimization, explainable AI, and immersive stakeholder engagement.

\subsection{Traditional ML Methods for Traffic Control}

AI/ML models embedded in digital twins learn from historical and real-time data to refine predictions and control policies, while the digital twin provides a realistic environment linking decisions to simulated outcomes~\citep{Lai2024Reinforcement, Yao2023Advanced}.
Regression and classification algorithms provide foundational analysis of traffic dynamics; for example, regression-based simulators capture complex spatiotemporal dependencies and support what-if exploration for policy and congestion estimation \citep{Mostafi2022RegTraffic}. Linear regression has been shown to yield accurate volume estimates on low volume roadways when appropriate explanatory variables are used \citep{Yeboah2023Estimating}. 
Classification models including random forests and support vector machines (SVM) are known to be useful in classifying traffic states and detecting anomalies, while time series forecasting models like autoregressive integrated moving average (ARIMA) model forecast short-term flows, accident hot spots, and energy consumption, as demonstrated in sustainable management frameworks \citep{Nautiyal2025Analysisa, Louati2025Machine}.
Together, digital twins and AI/ML enable proactive adaptation by sensing contextual changes at intersections (e.g., sudden influxes or emergency vehicles) and dynamically adjusting signal phases or traveler advisories \citep{Liu2023Deep}. These capabilities are further strengthened by ongoing advances in AI/ML algorithms, significantly enhancing traffic management.

\subsection{Reinforcement Learning and Multi-Agent Coordination}

As part of proactive traffic management, RL \citep{Sutton2018Reinforcement} has emerged as a leading AI approach for adaptive traffic signal control. 
In an RL setup, traffic signal controllers are considered agents that monitor various traffic conditions, including queue lengths, waiting times, and current signal phases, enabling them to execute actions such as adjusting the signal lights. 
This agent-environment interaction aims to maximize a predefined reward, often tied to performance metrics like minimizing vehicular delays and improving overall traffic throughput~\citep{Mousavi2017Traffic, Wei2021Recent}.
Over time, the RL agent ``learns'' an optimal or near-optimal policy by trying different actions and receiving feedback in the form of reward signals. 
This approach enables self-optimization that can adjust to complex traffic patterns without being explicitly programmed for every scenario~\citep{Aslani2017Adaptive}.
Reward functions in the literature often reflect traffic performance metrics, such as vehicle waiting time, queue length, or throughput\citep{Egea2020Assessment}. 
For example, if long queues form on one approach, the RL agent might learn to extend the green light to dissipate the queue, as this yields a higher cumulative reward due to reduced delays. 
Based on this, the agent is trained using a simulation and over many simulation cycles, the agent refines its timing decisions to optimally respond to demand fluctuations.
Studies 
have demonstrated that RL controllers can outperform pre-timed or actuated controllers by reacting more adaptively to incidents and daily traffic variations~\citep{Lai2024Reinforcement}. 


In early implementations, simple RL algorithms like Q-learning were used to adjust a single intersection’s lights, showing significant reductions in delay and stops.
More recently, deep RL methods have been applied, where deep neural networks approximate the optimal policy or value function. 
This approach allows handling larger state spaces (for example, using camera images or data from multiple intersections). 
Robust RL policies trained in digital twin environments adapt to disruptions like lane closures or incidents, enhancing urban grid resilience and informing future resilience models~\citep{Bubicz2023Planninga}.
Overall, RL provides a flexible and powerful toolkit for adaptive signal control, though its field deployment requires robust simulation training (often using digital twins) and careful attention to safety and fairness~\citep{Louati2025Machine}. 

At the network level, MARL strategies are being pursued. 
One study optimized signals across a city grid by coordinating agents at each intersection, leading to smoother traffic progression. 
Another innovative paradigm, dubbed ``democratizing traffic control,'' trained multiple RL agents, each with a different objective (e.g., minimizing travel time or emissions), and then had them collectively vote on the final signal plan in real time~\citep{Korecki2024Democratizing}.

MARL extends RL by coordinating agents at intersections via shared rewards or messages to optimize total travel time or balanced queues~\citep{Louati2025Machine}.
Cooperative adaptable lanes dynamically reallocate lanes using multiple RL agents, cutting average waiting times by almost 50\% compared to static allocations \citep{Dubey2024Cooperative}.
Agents with distinct objectives vote on signal plans, balancing travel time, emissions, and safety in a participatory MARL framework \citep{Korecki2024Democratizing}.

\subsection{Deep Learning for Prediction, Perception, and Control}
Deep learning significantly contributes to adaptive traffic management, particularly by capturing complex spatial and temporal traffic patterns using CNNs, recurrent neural networks (RNNs), and GNNs~\citep{Ju2024Comprehensivea, Jiang2022Grapha}.

Deep learning methods~\citep{LeCun2015Deepa}, excel at capturing complex spatial and temporal patterns in traffic data~\citep{Scarselli2009Grapha}.  
Hybrid models integrating YOLOv4  \citep{Bochkovskiy2020YOLOv4} detection with LSTM layers predict time incremental flows, enabling preemptive signal adjustments with higher accuracy~\citep{Shepelev2023Hybrid}. 
Spatial–temporal GNNs encode road networks as graphs, preserving topology for superior flow prediction and control recommendations~\citep{Li2023Graphpowered}.
Generative spatial AI, such as foundation-based Large Flow Models, fills missing traffic data and forecasts multimodal urban dynamics within digital twins, improving data fidelity~\citep{Huang2025Generativea}.  
Cooperative perception fuses camera and LiDAR data across connected autonomous vehicles and infrastructure via CNN or graph-based models, enhancing object detection and trajectory forecasts for incident-aware control~\citep{Yu2022review}.
Meanwhile, 3D-Net uses monocular cameras to detect and track road users in 3D, rapidly flagging incidents for downstream control adjustments~\citep{Rezaei20233DNeta}. 
DL-driven predictions feed into adaptive loops, such as dynamic green extensions or diversion alerts, closing the feedback loop in the digital twin framework~\citep{Liu2023Deep}.

\subsection{AI/ML-Driven Edge-to-Cloud Architectures and Immersive Decision Interfaces}

AI/ML-enabled digital twins leverage an edge–fog–cloud continuum to satisfy strict latency and compute requirements for continuous model updates and real-time inference~\citep{Al-Dulaimy2024computinga}. 
Complementary Narrow-Band IoT–linked digital twins employ lightweight ML models at the network edge to maintain energy efficient, scalable sensor fusion for large-scale deployments~\citep{Dangana2024Digital}.  
Meanwhile, economic–environmental frameworks automatically partition AI workloads to allocate preprocessing, inference, and training tasks between edge nodes and cloud servers, to optimize throughput, cost, and carbon footprint~\citep{Liu2024Automated}.  
Emerging quantum computing techniques promise to accelerate combinatorial optimization subroutines in AI traffic signal controllers, enabling sustainable, circular transport strategies~\citep{Jami2025Quantum}.  
Although originally developed for maritime operations, systems-driven methods integrate explainable AI modules within digital twins to generate operator-centric recommendations, validate safety constraints, and support human-in-the-loop tuning prior to field deployment \citep{Zhang2025Systems}.
Finally, gamified VR dashboards and metaverse interfaces harness interactive ML visualizations to engage planners and the public to allow immersive exploration, scenario analysis, and participatory tuning of twin-based control strategies~\citep{ReynosoVanderhorst2024Technologicala}.

\section{Safety and Protection of Vulnerable Road Users}
\label{sec:safety-vulnerable-users}

Building on the previous section’s discussion of how AI/ML integrated with digital twins can enhance traffic control, this section answers the question, \textit{``What methodologies enable a digital twin to detect, predict, and mitigate risks for vulnerable road users?''}


Vulnerable road users (VRUs), such as pedestrians and cyclists, face disproportionate risks with the highest crash rates per distance traveled and rising fatality trends in some regions \citep{Tengilimoglu2023Implications}. 
Enhancing digital twin models to better capture human behavior is critical, as it can reveal potential risks and enable proactive interventions to prevent crashes. 
For example, a recent work demonstrates how generative spatial AI can complete and predict flow data for pedestrians and vehicles alike, offering a powerful tool for pre-crash scenario analysis and VRU risk assessment~\citep{Huang2025Generativea} .
At a higher architectural level, a layered IoT–edge–cloud framework can support low-latency, distributed sensing and analytics for VRU protection in digital twins\citep{Al-Dulaimy2024computinga}. 

In total, this section reviews 15 papers that are related to vulnerable road user protection within digital twins.
The discussion is organized into three subsections: \textit{Sensing, Modeling, and Predictive Analytics}, which addresses data acquisition methods and AI/ML-based pedestrian behavior prediction; 
\textit{Behavioral and Infrastructure Simulations}, which examines simulation frameworks for VRU behavior and infrastructural interventions; 
and \textit{Ethical, Privacy, and Systems Perspectives}, which explores privacy concerns, bias mitigation, and system-level governance for safe and equitable digital twin deployments. 

\subsection{Sensing, Modeling, and Predictive Analytics}

Modern sensing techniques provide rich data on both road users and the surrounding environment. 
For example, unmanned aerial vehicles (UAVs) equipped with photogrammetry can rapidly survey intersections and generate detailed 3D models, identifying hazards such as foliage obstructing sight lines \citep{Congress2021Identifying}. 
On-vehicle multimodal sensor suites, combined with infrastructure-mounted sensors, ensure robust detection of pedestrians and cyclists by compensating for the limitations of any single sensor modality. 
A complementary approach is found in the zero calibration 3D object recognition system which uses only monocular cameras and GPS information to estimate 3D bounding boxes and trajectories for vehicles, cyclists, and pedestrians in real time \citep{Rezaei20233DNeta}. 
On the other hand, emergency vehicles with edge computing sensors to detect blind spot obstacles, including VRUs, at the vehicle perimeter, providing millisecond scale alerts to drivers~\citep{Mukhopadhyay2024Edge} . 
The deployment of roadside sensor networks, comprising of cameras, LiDAR, millimeter-wave radar, and weather stations, for cooperative intelligent driving has been shown to form the backbone of intelligent intersections by enabling global scene sensing and collaborative VRU detection in digital twin environments \citep{An2023Roadsidea}. 

Machine learning models learn pedestrian crossing behaviors or cyclist speed profiles from data, improving the prediction of potential conflict points. 
In vehicle-to-digital-twin systems, real-time AI-driven analytics can flag anomalous behavior (e.g., a pedestrian jaywalking) and assess collision risks. 
Notably, while camera-only systems detect less than 30\% of potential pedestrian collisions, fused sensor systems (integrating cameras, LiDAR, and radar) can prevent over 90\% of them \citep{Tengilimoglu2023Implications}. 
Furthermore, these predictive capabilities are also extended by enabling flow completion in unobserved locations and forecasting the evolution of VRU movements under different infrastructure scenarios~\citep{Huang2025Generativea} . 
Acoustic signal-based detection systems further augment these digital twins, a multi-criterion asynchronous classifier is used to detect accident hot spots and provide early alerts for VRUs, outperforming traditional fixed detector methods in accuracy and responsiveness~\citep{Yadav2025Leasta}.

\subsection{Behavioral and Infrastructure Simulations}

Several methodologies have been proposed to capture and predict VRU behavior within traffic digital twins. 
One approach integrates continuous real-time calibration of microscopic simulators, such as SUMO, with live sensor data, enabling digital twins to dynamically adapt to changing traffic conditions and pedestrian movements \citep{Kusic2023digitala}.
Another methodology uses agent-based pedestrian network digital twins, which focus on discrete network graphs to simulate individual pedestrian rerouting behaviors, effectively capturing the stochastic nature of route choice \citep{Jabbari2023Pedestrian}. 
MARL has also been used to dynamically reallocate lane widths and right-of-way permissions, significantly reducing queue lengths and enhancing safety in mixed traffic scenarios \citep{Dubey2024Cooperative}. 
Meanwhile, probabilistic pedestrian crossing models combined with Monte Carlo simulations provide insights into pedestrian risk behaviors under dynamic traffic conditions, emphasizing critical safety thresholds related to vehicle proximity and traffic signal phases \citep{Bagabaldo2025Improving}.
Additionally, infrastructure simulations underscore the importance of modeling interactions between autonomous vehicle-centric infrastructure changes and VRU movements to ensure safety in mixed traffic environments \citep{Tengilimoglu2023Implications}.

Digital twins also incorporate detailed road infrastructure attributes, ranging from intersection geometries to signal phasing and crosswalk types. 
Simulation frameworks can test modifications, such as converting an unsignalized crosswalk to a signalized one or adding a pedestrian refuge island, while remote sensing and GIS-based methods (e.g., high-resolution 3D digital surface models from UAV surveys) simulate drivers’ and pedestrians’ sight lines \citep{Congress2021Identifying}. 
Combining spatial crash models (e.g., Geographically Weighted Poisson Regression, which is a spatial regression model for count processes wiht local variation, with macroscopic statistical models (e.g., Negative Binomial) within a digital twin architecture improves hot spot identification accuracy by over 10\% relative to single model approaches \citep{Rua2024Combinationa}.

\subsection{Ethical, Privacy, and Systems Perspectives}

Eventually, human behavior in traffic is influenced by social, cultural, and ethical factors.
Methodologies incorporating expert knowledge and human factors are vital, ensuring that digital twins respect privacy, avoid algorithmic discrimination, and support socially acceptable interventions. 
Delphi studies, which is a structured, iterative survey technique that uses multiple rounds of anonymized questionnaires and controlled feedback among a panel of experts, and multi-stakeholder workshops have identified key concerns such as privacy, bias, and safety when introducing AI-driven technologies \citep{Stahl2023Exploring}. 
The ethical and privacy implications of pervasive VRU sensing are highlighted, especially the need for robust anonymization and informed consent when aggregating individual movement data~\citep{Fadhel2024Comprehensivea}.
GeoAI frameworks further emphasize values-driven co-design, ensuring that digital twin interventions align with community values and equitable access goals~\citep{Mortaheb2023Smarta}.

\section{Scaling from Localized Intersections to Citywide Traffic Networks}
\label{sec:scaling-int}

In the previous sections, we established the key components of an ideal digital twins for intelligent intersection.  
It is worth noting, however, that many existing digital twin implementations have primarily focused on localized or single intersections~\citep{Kampourakis2023systematic, Wang2024Architecture}.  
Extending these applications to encompass entire citywide traffic networks remains a significant challenge \citep{Kusic2023digitala} and requires tackling issues related to computational efficiency, distributed control, and secure data communication.  
This section presents a synthesis of 35 papers on digital twin applications in smart infrastructure, with a focus on scaling from localized intersections to citywide traffic networks. The discussion is organized into three subsections: \textit{Digital Twin Applications Across Smart and Urban Infrastructure}, which examines foundational concepts from CPS, Industry 4.0, and urban computing, as well as the integration of AI, IoT, and spatial data models; \textit{Control and Communication Frameworks for Scalability}, which explores distributed control architectures, semantic interoperability, and secure communication protocols such as V2X and blockchain; and \textit{Challenges in Scaling Digital Twin Applications}, which highlights key computational, organizational, and technological hurdles and identifies promising strategies such as lightweight models, parallel intelligence, and secure wireless systems for enabling large-scale deployment.

\subsection{Digital Twin Applications Across Smart and Urban Infrastructure}

Example smart infrastructure digital twin use cases, from traffic control to smart mobility planning are presented in Table \ref{tab:dt-applications} in the appendices.
Related literature also provides a comprehensive overview of digital twin concepts within CPS and Industry 4.0 environments.  
For instance, a systematic review of wireless security testbeds in CPS establishes the groundwork for understanding the convergence of physical and digital realms~\citep{Kampourakis2023systematic}.  
Complementary investigations into urban computing and building information modeling for transportation infrastructure emphasize the integration of digital models with physical structures, a critical prerequisite for scalable digital twin applications~\citep{Alshuwaikhat2022Analysis, Pirdavani2023Application}.  
Research on digital twin applications in Industry 4.0 and intelligent highway infrastructure further demonstrates the potential benefits of digital replication in complex urban environments~\citep{Wang2024Architecture, Javaid2023Digital}.  
Moreover, studies on blockchain, IoT, and AI in logistics emphasize the importance of secure, distributed data management~\citep{Idrissi2024Blockchain}.  
Foundational paradigms for digital twins adoption in the construction sector underscore its transformative potential beyond transportation~\citep{Moshood2024Infrastructurea}.  
Localized digital twin implementations aimed at urban inclusivity have also emerged, showcasing modular approaches for age-friendly environments~\citep{Villanueva-Merino2024Leveraging}.  
A systematic review highlights the integration of AI within digital twin systems across various domains~\citep{Kreuzer2024Artificiala}.

Additional studies expand upon urban digital twin implementations by focusing on efficient traffic data collection and computational processing.  
Research on traffic organization and safety highlights the necessity for effective data collection~\citep{Pugachev2020Features}.  
Advances in generative spatial AI for sustainable smart cities and pioneering large flow models for urban digital twin frameworks~\citep{Huang2025Generativea} offer promising methods for processing and simulating complex urban flows.  
Analyses of Geo CPS with a focus on spatial challenges further illustrate opportunities and constraints in geospatial data integration~\citep{Yan2019Geo}.  
Comprehensive methodologies for fusing heterogeneous urban data sources have also been reviewed~\citep{Fadhel2024Comprehensivea}.  
Investigations into sustainable integration of digital twin in road infrastructure~\citep{Ulrich2023Howa} and the impact of connected corridor volume data imputations on digital twin performance~\citep{Saroj2023Impacta} directly address challenges related to data imputation and real-time performance, while studies on wireless network digital twins for smart railways~\citep{Guan2024Key} and methods for lightweighting digital twin information models~\citep{Jin2024Lightweighting} provide practical approaches to reducing computational loads.  
Dedicated investigations into lightweighting processes for digital twin information models demonstrate strategies for efficient data handling in smart city services.

\subsection{Control and Communication Frameworks for Scalability}

A significant portion of the literature focuses on the control and communication frameworks essential for digital twin scalability.  
Insights into evaluating performance of measurement equipment in automated traffic control systems~\citep{Safiullin2020Method} and the design of multi-level traffic management systems~\citep{Kerimov2020Model} contribute valuable perspectives on distributed control architectures.  
High performance computing platforms have also been leveraged for distributed route computations in traffic flow models~\citep{Silva2025HighPerformance}.  
Agent-based digital twin architectures have been demonstrated in logistics applications to integrate multiple stakeholders and assets~\citep{Xu2025Multiagent}.  
Semantic interoperability frameworks aligning BIM and IoT sensor data models have been proposed to ensure consistent data exchange across digital twin and BIM systems~\citep{Okonta2024Semantic}.  
Further studies on organizational tensions in Industry 4.0 reveal managerial and structural challenges for integrating digital twin systems~\citep{Dieste2022Organizational}.  
Testing range principles for intelligent transportation technologies~\citep{Zhankaziev2020Principles} and bibliometric analyses quantifying the evolution of IoT within smart infrastructure~\citep{Stan2024Quantifying} highlight the need for robust and scalable data infrastructures.  
Investigations into roadside sensor network deployments~\citep{An2023Roadsidea} and GeoAI applications for re-imagining smart cities~\citep{Mortaheb2023Smarta} further stress the importance of distributed sensing frameworks.  
Advanced V2X communication and digital twin integration for traffic light systems~\citep{Wagner2023SPaTa} illustrate secure, low-latency data exchange, while technological visions for Meta Smart Twin Cities emphasize the convergence of metaverse and digital twin paradigms~\citep{Mogaji2023Metaverse}.  
Exploratory studies have investigated quantum computing to optimize large-scale digital twin simulations and circular economy models, pointing to significant performance gains for complex urban networks~\citep{Jami2025Quantum}.  
Meanwhile, non-fungible token (NFT) mechanisms have been proposed to provide secure provenance and asset management for distributed digital twin ecosystems~\citep{Davies2024Nonfungible}.  
NFT is a cryptographically unique, indivisible, irreplaceable and verifiable token implemented via blockchain smart‑contract protocols that enables the creation, ownership, transfer and verification of unique digital (or physical) assets \citep{Valeonti2021Crypto}.
Democratized control approaches involving multiple stakeholders and learning-based controllers in voting-based signal management have also been proposed~\citep{Korecki2024Democratizing}.  
Finally, the synergistic interplay between AI and digital twins in environmental planning~\citep{Bibri2024synergistic} and interdisciplinary approaches such as parallel intelligence~\citep{Zhao2023parallel} suggest integrative strategies to overcome challenges in both computation and control.

\subsection{Challenges in Scaling Digital Twin Applications}

Scaling digital twins applications from localized intersections to citywide networks presents several critical challenges.  
Efficient processing of vast urban datasets is essential for real-time simulation and control; however, integrating complex 3D models with continuous sensor feeds remains computationally intensive~\citep{Pirdavani2023Application}.  
Although approaches such as lightweight digital twin models~\citep{Jin2024Lightweighting} and advanced generative spatial AI frameworks~\citep{Huang2025Generativea} have been proposed to mitigate computational demands, these challenges persist.  
Moreover, managing digital twins across extensive geographical areas requires the development of distributed control architectures capable of coordinating data from numerous intersections and road segments.  
While multi-level system models~\citep{Kerimov2020Model} and performance evaluation methodologies~\citep{Safiullin2020Method} provide promising strategies, they also highlight the complexities inherent in resolving managerial and structural issues.

Additionally, for citywide applications, ensuring secure and reliable data communication is paramount.  
Models based on blockchain, IoT, and AI in logistics~\citep{Idrissi2024Blockchain} and advanced V2X communication systems~\citep{Wagner2023SPaTa} are essential for maintaining data integrity and protecting against cyber threats across distributed networks.  
Organizational challenges in Industry 4.0 implementations further complicate integration efforts, as described in studies on structural and managerial tensions~\citep{Dieste2022Organizational}.

The literature indicates that several key technological advancements are necessary to overcome these challenges.  
Enhanced computational algorithms and lightweight digital twin models are critical; the development of efficient algorithms, leveraging approaches such as generative spatial AI~\citep{Huang2025Generativea} and parallel intelligence~\citep{Zhao2023parallel}, can significantly improve computational efficiency, while techniques for reducing the data footprint of digital twin information models~\citep{Jin2024Lightweighting} help maintain essential detail without overwhelming system resources.  
Equally important is the need for robust distributed control and hierarchical management frameworks.  
Distributed, multi-level control architectures~\citep{Kerimov2020Model} that integrate localized controllers with central supervisory systems offer flexibility and resilience, yet the successful implementation of such systems also requires addressing inherent organizational tensions~\citep{Dieste2022Organizational}.  
Furthermore, establishing secure and reliable data communication protocols is critical.  
The adoption of next generation wireless technologies, including 5G/6G and V2X protocols~\citep{Wagner2023SPaTa, Gallego-Madrid2023role}, combined with blockchain-based solutions and robust encryption methods~\citep{Idrissi2024Blockchain}, is essential to ensure low-latency and secure data exchange across widespread, distributed networks.

\section{Discussion}
\label{sec:discussion}

Our comprehensive literature review reveals a rapidly advancing landscape in the application of digital twins for transportation that supports intelligent intersections. The research landscape, categorized into five key themes, demonstrates significant progress in developing sophisticated architectures, leveraging advanced data processing and simulation, integrating AI/ML for adaptive control, safety and protection of VRUs, and scaling localized intersection to network-level for cities. However, this review also illuminates critical gaps and unchallenged assumptions that temper the current optimism. This discussion synthesizes these findings, offers a critical perspective on the prevailing research trajectory, and outlines a path forward for developing digital twins that are not merely efficient, but fundamentally safe and human-centric.

\subsection{The Simulation-Reality Gap: From Virtual Models to True Digital Twins}
A recurring theme throughout the reviewed literature is the heavy reliance on simulation. While simulators like SUMO and co-simulation platforms coupling traffic and vehicle models (e.g., CARLA) are invaluable for initial development and testing, a significant portion of the reviewed literature treats the ``digital twin'' as a sophisticated, yet one-way, simulation. Data flows from the physical world (or a proxy for it) into the model, which then runs scenarios and proposes optimizations. This falls short of the true digital twin paradigm and may just fall into what is called a ``digital shadow,'' which is a virtual representation that automatically mirrors real‑time data from a physical system, but lacks the ability to influence or control it~\citep{Kritzinger2018Digital}.

The essence of a digital twin, as envisioned by \citet{Grieves2014DigitalTwin} and NASA \citep{Glaessgen2012Digitala}, lies in its continuous, bidirectional synchronization with its physical counterpart. The digital model should not only reflect the state of the physical intersection but also be capable of actuating changes whose effects are then measured and fed back into the model, closing the loop. Most studies, however, stop at the simulation stage, presenting performance improvements in a virtual environment without grappling with the complexities of real-world deployment and feedback or separately training AI/ML models for traffic control which later on applied but does not affect traffic in real time. Without this feedback loop, the digital model risks becoming a ``digital shadow,'' reflecting a reality that has already passed and making decisions based on potentially stale or idealized data. This one-way data flow fails to account for the inherent uncertainty and unpredictability of the physical world, from sensor noise and communication latency to unexpected human behavior.

To fully realize the transformative potential of digital twins for intelligent intersections, we must move beyond simulation and passive monitoring toward real-time, closed-loop systems that can actively implement, test, and adapt control strategies in the physical world. This shift from virtual analysis to actionable intervention is essential to ensure that digital twins are not just analytical tools but operational systems that improve safety, efficiency, and responsiveness on the ground.

\subsection{The Human Blind Spot: Prioritizing Pedestrians and Vulnerable Road Users} 

Our review reveals a pronounced vehicular-centric bias in the current body of research. The primary metrics for success are almost universally tied to traffic flow: reducing vehicle delay, minimizing queue lengths, and increasing throughput. While these are important goals, they often overshadow the most critical function of an intersection, which is to ensure the safety of all users, especially the most vulnerable.

VRUs, such as pedestrians, cyclists, and micromobility users, are frequently treated as an afterthought or are modeled with simplistic assumptions. While vehicles adhere to predictable kinematic models, pedestrians and cyclists exhibit complex, stochastic behaviors influenced by social cues, perceived risk, and environmental context. The review shows a dearth of research integrating fine-grained, agent-based models of pedestrian behavior into digital twins. Most models either ignore pedestrians or use coarse, aggregated flow approximations. This is a critical oversight. An intersection that is highly efficient for cars but terrifying or dangerous for a child, an elderly person, or a person with a disability has failed in its fundamental civic duty. Thus, agencies like USDOT have initiatives and funding dedicated to safer roads, including the Safe Streets and Roads for All (SS4A) competitive grant program \citep{USDOT2025SS4A} and the U.S. DOT Research, Development, and Technology Strategic Plan for Fiscal Years 2022–2026, which outlines the Department’s research priorities, objectives, and strategies for improving transportation safety, equity, climate resilience, and innovation \citep{USDOT2022RDTSPlan}. We envision that a true digital twin balances traffic flow efficiency with pedestrian safety, going beyond mere automation to support more holistic and inclusive mobility solutions. Our focus on safety-first digital twins seeks to rebalance these priorities, ensuring that VRU protection is a primary objective function, not a secondary constraint.

\subsection{The Ethics of Surveillance, Bias, and Accountability in Digital Twins} 

As digital twins evolve from passive simulations to active, decision-making systems embedded in urban infrastructure, their ethical implications demand careful scrutiny. Despite increasing sophistication in data-driven control algorithms and sensor-rich environments, the reviewed literature rarely interrogates the full spectrum of ethical concerns surrounding their real-world deployment.

First, there are serious privacy concerns. Digital twins for intelligent intersections often rely on high-resolution sensor data, such as video feeds, LiDAR, and mobile device signals, to capture real-time traffic dynamics. While essential for accurate modeling and control, such data streams can inadvertently capture personally identifiable information, especially of pedestrians and cyclists. Without strong safeguards, such as robust anonymization protocols, secure data handling, and transparent data governance, these systems risk enabling mass surveillance, eroding public trust and infringing on privacy rights.

Second, algorithmic bias poses a critical risk. Models trained on biased datasets due to under-representation of certain populations, skewed spatial sampling, or historical inequities, can reinforce and even amplify systemic disparities. For example, optimization focused primarily on vehicle throughput in affluent areas may deprioritize pedestrian safety or service levels in underserved neighborhoods. Ensuring fairness requires embedding ethical frameworks and fairness-aware learning mechanisms into model design and training pipelines from the outset.

Third, reliability and misuse must be addressed. A system as powerful as a digital twin can be exploited, either intentionally or unintentionally, by its human operators, or manipulated through cyber-physical vulnerabilities. Safeguards must be implemented to prevent tampering, adversarial input manipulation, or human misuse of override functions. Moreover, the digital twin must be resilient against failures in the physical system or unexpected real-world events, and should be designed to fail safely, with clearly defined human-in-the-loop override protocols. Ensuring reliability is not just a matter of technical robustness but also of institutional checks and balances.

Finally, questions of transparency and accountability remain unresolved. As digital twins increasingly influence real-world decisions, such as modifying signal timing or prioritizing specific user flows, mechanisms for oversight and public communication become essential. Questions arise such as: \textit{``Who is responsible if an AI-driven decision results in harm?'', ``How can these decisions be understood and contested by non-experts?'',} and \textit{``What transparency guarantees exist to prevent misuse or unchecked authority?''}.

These are not purely technical challenges, but socio-technical ones that demand interdisciplinary solutions across engineering, computer science, ethics, law, and urban governance. Addressing them requires collaboration among a diverse array of stakeholders, including municipal and regional governments, national regulatory agencies, urban planners, civil society organizations, academic institutions, private sector technology developers, and the general public, whose mobility, privacy, and safety are most directly affected. Ethical integrity, reliability, and resistance to misuse must be foundational components of any digital twin system intended for public deployment.

\subsection{The Challenge of Validation and Sparse Historical Data}

The review also raises two pragmatic yet profound questions that are largely unaddressed:
\begin{enumerate}
    \item \textit{How can the effectiveness of a new traffic management strategy be rigorously and safely tested before live deployment?}
    \item \textit{What happens when historical data is sparse or non-existent for a particular intersection or a rare event?}
\end{enumerate}

Regarding the first question, a digital twin offers a revolutionary framework for validation that goes beyond pure simulation. We propose a multi-stage validation process: \textit{Simulation-in-the-Loop (SIL):} The initial phase, where control logic is tested in a fully virtual environment. \textit{HIL:} The control algorithms are run on the actual traffic controller hardware, which interacts with the simulated digital twin. \textit{Shadow Mode Deployment:} This is the crucial, and often missing, step. The digital twin is deployed in the field, receiving live sensor data and making decisions in parallel with the existing intersection controller. However, its decisions are \textit{not} actuated. Instead, they are logged and compared against the real-world outcomes. This allows for the validation of the twin's performance and safety under real-world conditions without any risk to the public. \textit{Phased Live Deployment:} Only after passing rigorous shadow mode validation can the system be gradually brought online, perhaps during off-peak hours and with human oversight and safety overrides.

Another point is the absence of training data, digital twins must incorporate methods beyond traditional supervised learning. For new intersections or for predicting rare but catastrophic events like near-misses or crashes, there is no `big data' to learn from. Here, the future lies in probabilistic models and combining physics-based models with modern AI. Generative AI techniques (e.g., GANs, diffusion models) can be used within the digital twin to create vast libraries of synthetic but realistic ``edge case'' scenarios to train and harden the control logic. Furthermore, principles of transfer learning can be applied, where a model trained on a data-rich intersection can be fine-tuned with minimal data for a new location.

While the field has made remarkable strides in building the components of digital twins for intelligent intersections, our review suggests that the current trajectory is focused on computational sophistication rather than holistic, real-world impact. The next frontier is to bridge the simulation-reality gap, shift from a vehicle-centric to a human-centric paradigm, and develop robust frameworks for validation and learning under uncertainty. 

We can see that future research can be positioned to address these gaps, aiming to develop digital twins that are not just computationally powerful, but are fundamentally architected for safety, human-centricity, and real-world fidelity.

\section{Conclusion}
\label{sec:conclusion}

This review shows how digital twin technology offers a transformative framework for managing and optimizing intelligent intersections. By synthesizing findings across five major thematic areas, we capture the state of the art, key challenges, and emerging directions for research and implementation. The major takeaways are summarized as follows:

\begin{enumerate}
    \item \textit{Digital Twin Architectures and Frameworks:}  
    Recent literature highlights the emergence of modular, multilayered digital twin architectures that incorporate edge computing, microservice-based platforms, and AI-enhanced components. These architectures typically include data acquisition, communication, processing, modeling, and visualization layers, all coordinated through cloud native frameworks. Interoperability and standardization via open APIs, microservices, and semantic models remain critical for integrating diverse sensor systems and simulation tools. These advancements address the pressing need for scalable and flexible digital twin systems that can evolve alongside technological innovations.

    \item \textit{Data Processing and Simulation Techniques:}  
    Effective digital twins depend on robust real-time data ingestion and heterogeneous fusion across multiple sensor modalities (e.g., LiDAR, radar, loop detectors). Studies increasingly apply probabilistic methods (e.g., Kalman filtering, clustering, and Bayesian learning) and leverage AI for imputation and anomaly detection. Simulation platforms like SUMO, CARLA, and co-simulation frameworks are widely used, often augmented by synthetic data generation and AI-enhanced virtual perception modules. These tools enable high fidelity testing and calibration, supporting predictive and adaptive traffic management despite data gaps or uncertainty.

    \item \textit{AI/ML in Traffic Control:}  
    AI/ML algorithms, particularly RL, DL, and GNNs, are at the core of adaptive traffic signal control in digital twins. RL agents can learn optimal signal timings through simulation, while multi-agent systems coordinate control across intersections. Deep learning models support traffic prediction, anomaly detection, and perception fusion. Increasingly, these models operate within edge-to-cloud environments, allowing real-time inference and continuous model refinement. Some implementations even explore participatory or explainable AI frameworks to support fairness, transparency, and collaborative decision-making.

    \item \textit{Safety and Protection of Vulnerable Road Users:}  
    Emerging approaches integrate explainable AI, social force models, and agent-based simulations to model pedestrian and cyclist behaviors. VRU detection and protection are enhanced via multimodal sensing (e.g., LiDAR, mmWave radar), zero calibration vision systems, and real-time fusion analytics. Simulation platforms increasingly incorporate behavioral diversity and infrastructure variation to evaluate how changes affect VRU risk. Ethical and privacy concerns, such as data anonymization and inclusive design, are also being integrated into digital twin development, ensuring human-centric safety interventions.

    \item \textit{Scaling to Citywide Traffic Networks:}  
    Moving from intersection-level digital twins to citywide implementations requires distributed architectures, lightweight modeling strategies, and secure, low latency communication protocols. Key enablers include 5G/6G networks, blockchain-based data verification, and edge-cloud orchestration frameworks. Recent advances demonstrate the feasibility of coordinated control across corridors, dynamic routing, and integration with smart infrastructure. However, significant challenges persist in terms of computational efficiency, standardization, and organizational coordination. Addressing these issues will be essential for realizing the full potential of digital twins in sustainable and intelligent urban mobility systems.
\end{enumerate}

Despite impressive technological advancements, this review reveals a critical disconnect between the growing computational capabilities of digital twins and their deployment as safe, ethical, and human-centered systems. Four foundational challenges undermine the current trajectory.

First, the persistence of the \textit{simulation-reality gap} means many so-called digital twins remain as unidirectional simulations, which is referred to in literature as ``digital shadows,'' where they are incapable of acting upon or learning from the physical world. Without real-time, bidirectional feedback loops and closed-loop deployment protocols, these systems cannot fulfill their promise of dynamic adaptation and robust control.

Second, there remains a pronounced \textit{human blind spot}. The vast majority of research prioritizes vehicle throughput, often at the expense of pedestrians, cyclists, and other vulnerable road users. True intelligence in traffic systems requires not just optimizing flows, but safeguarding all road users, particularly those most at risk. Rebalancing this focus is essential to align digital twins with broader societal goals such as equity, inclusivity, and safety.

Third, the field has not adequately addressed the \textit{ethical and governance challenges} posed by real-world deployment. As digital twins evolve from passive simulations to active decision-making systems, issues such as privacy, algorithmic bias, and accountability become critical. Without strong safeguards for anonymization, fairness-aware optimization, and transparent oversight, these systems risk becoming instruments of surveillance or unintended discrimination. Interdisciplinary collaboration and ethical frameworks must be embedded into both technical design and institutional deployment.

Fourth, robust mechanisms for \textit{validation and learning in data scarce environments} are still lacking. Many studies assume rich historical data and idealized conditions. However, the real-world deployment of digital twins demands rigorous testing under uncertainty, from shadow mode validation to the integration of synthetic data, transfer learning, and probabilistic reasoning for edge case prediction.

These lead us to conclude that future research must move decisively from one-way analysis to active implementation. This includes embedding human-in-the-loop safeguards, developing equitable optimization objectives, and establishing standards for ethical deployment. The next generation of digital twins should not merely reflect traffic conditions but they must actively shape safer, more inclusive, and responsive urban environments.

In summary, the path forward lies in reimagining digital twins not as isolated technical constructs, but as socio-technical systems embedded in the lived fabric of our cities. Only by bridging the simulation-reality divide, centering human values, and institutionalizing ethical rigor can we realize the full potential of digital twins to transform intersections from bottlenecks into intelligent, inclusive, and adaptive components of future mobility systems.

\section*{Declaration of Generative AI and AI-assisted technologies in the writing process}
During the preparation of this work, the authors used \textit{Gemini 1.5 Pro} and \textit{2.0 Pro}, generative AI technologies, to expedite the filtering of relevant articles and sorting them across different themes, synthesizing of literature and refining the manuscript's language and structure. ChatGPT 4o and o4-mini-high provided editorial suggestions to enhance clarity, coherence, and consistency, and these AI-assisted outputs were then critically reviewed and meticulously revised by the authors who assume full responsibility for the intellectual content and accuracy of the final publication.

\section*{Declaration of competing interest}
The authors declare that they have no known competing financial interests or personal relationships that could have appeared to influence the work reported in this paper.

\section*{Acknowledgement}
The authors would like to thank the Princeton University School of Engineering and Applied Science (SEAS) Innovation Grant for funding this work.

\appendix

\section*{Appendix A: Acronyms and Abbreviations}
\setlength\emergencystretch{2em}
Table \ref{tab:abbreviations} provides the list of acronyms and abbreviations used in this paper.

\setcounter{table}{0}
\renewcommand{\thetable}{A.\arabic{table}}

\begin{small}
\begin{longtable}{|p{2.6cm}|p{4.2cm}|p{8.2cm}|}
\hline
\textbf{Acronym} & \textbf{Term} & \textbf{Brief Description} \\ \hline
AI/ML & Artificial Intelligence/Machine Learning & Computational methods that perform tasks requiring human-like intelligence. \\ \hline
ANN & Artificial Neural Network & ML model composed of interconnected nodes for pattern learning and prediction. \\ \hline
API & Application Programming Interface & Interface enabling software components to communicate. \\ \hline
ARIMA & Autoregressive Integrated Moving Average & Time-series model for short-term forecasting. \\ \hline
AV & Autonomous Vehicle & Vehicle with automated perception, planning, and control. \\ \hline
BIM & Building Information Modeling & Digital representation of built assets for visualization/management. \\ \hline
C\,-V2X & Cellular Vehicle-to-Everything & Cellular standard enabling V2X communications. \\ \hline
CNN & Convolutional Neural Network & Deep learning model effective for image/video tasks. \\ \hline
CPS & Cyber-Physical Systems & Integrated computational and physical processes. \\ \hline
DBSCAN & Density-Based Spatial Clustering of Applications with Noise & Unsupervised clustering algorithm robust to noise. \\ \hline
DL & Deep Learning & Type of machine learning that uses artificial neural networks with multiple layers to analyze data and learn complex patterns\\ \hline
DT & Digital Twin & Virtual counterpart of a physical system with live data sync. \\ \hline
DTA & Dynamic Traffic Assignment & Traffic modeling framework for route/flow evolution. \\ \hline
DSRC & Dedicated Short-Range Communications & Short-range wireless standard for vehicular comms. \\ \hline
EM & Expectation–Maximization & Iterative algorithm for latent variable parameter estimation. \\ \hline
GAN & Generative Adversarial Network & Generative deep learning framework (generator vs discriminator). \\ \hline
GMM & Gaussian Mixture Model & Probabilistic model representing data as a mixture of Gaussians. \\ \hline
GNN & Graph Neural Network & Deep learning on graph-structured data. \\ \hline
HIL & Hardware-in-the-Loop & Real-time coupling of controllers with simulated environments. \\ \hline
IoT & Internet of Things & Network of connected sensors/devices. \\ \hline
LiDAR & Light Detection and Ranging & Active remote-sensing technology for range/3D perception. \\ \hline
LLM & Large Language Model & Foundation model trained on large corpora for language tasks. \\ \hline
MARL & Multi-Agent Reinforcement Learning & RL with multiple interacting/coordinated agents. \\ \hline
mmWave & Millimeter Wave & High-frequency band supporting high-throughput links. \\ \hline
MQTT & Message Queuing Telemetry Transport & Lightweight publish/subscribe messaging protocol. \\ \hline
OMNeT++ & Objective Modular Network Testbed in C++ & Discrete-event simulation framework for networks. \\ \hline
ODD & Operational Design Domain & Conditions under which an automated system is designed to operate. \\ \hline
PCA & Principal Component Analysis & Dimensionality-reduction/feature extraction method. \\ \hline
PD & Proportional–Derivative & Feedback control law using proportional and derivative terms. \\ \hline
PLC & Programmable Logic Controller & Industrial controller used in real-time control. \\ \hline
Redis & Remote Dictionary Server & In-memory data store for caching/queues/pub-sub. \\ \hline
RL & Reinforcement Learning & Learning control via reward-driven interaction. \\ \hline
RSU & Roadside Unit & Fixed V2X communication node at roadside. \\ \hline
SPaT/MAP & Signal Phase and Timing / MAP as intersection geometry & V2X messages conveying signal timing and intersection geometry. \\ \hline
STGAT & Spatio-Temporal Graph Attention Network & GNN variant modeling spatial-temporal dependencies with attention. \\ \hline
SUMO & Simulation of Urban MObility & Microscopic traffic simulation platform. \\ \hline
TraCI & Traffic Control Interface & API coupling SUMO with external controllers. \\ \hline
UAV & Unmanned Aerial Vehicle & Aircraft without onboard human pilot. \\ \hline
V2X & Vehicle-to-Everything & Communication among vehicles, infrastructure, devices, etc. \\ \hline
VAE & Variational Autoencoder & Probabilistic generative model for representation learning. \\ \hline
VRU & Vulnerable Road User & Non-motorists such as pedestrians and cyclists (and similar users). \\ \hline
5G NR & Fifth-Generation New Radio & 5G air-interface standard supporting C-V2X and low-latency links. \\ \hline
\caption{Acronyms and Abbreviations Used in the Paper}
\label{tab:abbreviations}
\end{longtable}
\end{small}

\section*{Appendix B: Comparison of Digital Twins Review Papers with Respect to Intersection Focus}
\setcounter{table}{0}
\renewcommand{\thetable}{B.\arabic{table}}
\setlength\emergencystretch{2em}

\setlength\LTleft{0pt}
\setlength\LTright{0pt}

\setcounter{table}{0}
\renewcommand{\thetable}{B.\arabic{table}}
\setlength\emergencystretch{2em}

This appendix presents a comparative overview of existing literature reviews on digital twins, highlighting each paper’s main contributions, the research fields they cover, and the gaps or future work they identify.

\setlength\LTleft{0pt}
\setlength\LTright{0pt}

\begin{small}

\begin{longtable}{|>{\raggedright\arraybackslash}m{1.5in}|c|>{\raggedright\arraybackslash}m{1.45in}|>{\raggedright\arraybackslash}m{1in}|>{\raggedright\arraybackslash}m{1.45in}|}
  \caption{Comparison of Digital Twins review papers with respect to intersection focus.} \label{table:comparison-dt-reviews} \\
  \hline
  \textbf{Title of paper} & 
  \textbf{Year} &
  \textbf{Contributions} &
  \textbf{Field(s)} &
  \textbf{Gaps / Future Work} \\
  \hline
  \endfirsthead

  \multicolumn{5}{c}{{\bfseries \tablename\ \thetable{} -- continued from previous page}} \\
  \hline
  \textbf{Title of paper} & 
  \textbf{Year} &
  \textbf{Contributions} &
  \textbf{Field(s)} &
  \textbf{Gaps / Future Work} \\
  \hline
  \endhead

  \hline \multicolumn{5}{r}{{Continued on next page}} \\ 
  \endfoot

  \hline
  \endlastfoot

  Digital Twins for cities: Analyzing the gap between concepts and current implementations with a specific focus on data integration \cite{Jeddoub2023Digitalb} &
  2023 &
  \begin{itemize}
    \item Reviewed DT definitions vs.\ 3DCM, CIM, SDI
    \item Proposed three‐level data‐integration maturity model: schema, database, application
  \end{itemize} &
  Urban geospatial; Smart Cities; AEC &
  \begin{itemize}
    \item No unified DT definition
    \item Lack of generic data integration framework
  \end{itemize} \\
  \hline

  The Hitchhiker’s Guide to Fused Twins: A Review of Access to Digital Twins In Situ in Smart Cities \cite{Grubel2022Hitchhikers} &
  2022 &
  \begin{itemize}
    \item Defined ``Fused Twins'' combining DT + Situated Analytics
    \item Classified emerging prototypes and five DT components
  \end{itemize} &
  Smart Cities; Remote Sensing; AR/VR; Immersive Analytics &
  \begin{itemize}
    \item Need for embodied, in‐situ interfaces
    \item Evaluate cognitive benefits in real deployments
  \end{itemize} \\
  \hline

  City Digital Twin Potentials: A Review and Research Agenda \cite{Shahat2021Citya} &
  2021 &
  \begin{itemize}
    \item Mapped current/prospective DT city benefits and challenges
    \item Proposed research agenda: data efficiency, socio‐economic integration, twin coupling
  \end{itemize} &
  Sustainability; Smart Cities; Urban management &
  \begin{itemize}
    \item Early‐stage field; missing socioeconomic modules
    \item Need full digital-physical integration frameworks
  \end{itemize}  \\
  \hline

  Exploring Digital Twin Adaptation to the Urban Environment: Comparison with CIM to Avoid Silo-based Approaches \cite{Depretre2022EXPLORINGa} &
  2022 &
  \begin{itemize}
    \item Compared DT vs.\ CIM via literature review, practitioner survey and four case studies
    \item Highlighted governance and technical heterogeneity
  \end{itemize} &
  Urban planning; City modelling; CIM/DT frameworks &
  \begin{itemize}
    \item Fuzzy labels; siloed research streams
    \item Call for interdisciplinary, task-oriented typologies
  \end{itemize}  \\
  \hline

  Autonomous Driving Test Method Based on Digital Twin: A Survey \cite{Niaz2021Autonomousa} &
  2021 &
  \begin{itemize}
    \item Surveyed DT origins, CPS integration, industrial practices
    \item Proposed V2X-enabled DT test framework for connected vehicles
  \end{itemize} &
  Autonomous driving; CPS; V2X testing &
  \begin{itemize}
    \item DT definition ambiguity in transport
    \item Need real-world validation of V2X-DT testbeds
  \end{itemize}  \\
  \hline

  A comprehensive review of Digital Twin technologies in smart cities \cite{Huzzat2025comprehensivea} &
  2025 &
  \begin{itemize}
    \item Surveyed DT enabling tech (IoT, ML, CPS, blockchain)
    \item Reviewed urban case studies across domains and outlined challenges
  \end{itemize} &
  Smart Cities; Digital Engineering; Urban development &
  \begin{itemize}
    \item Lack of standard metrics and governance models
    \item Need cross‐domain, longitudinal deployments
  \end{itemize}  \\
  \hline

  AI‐Powered Digital Twins and Internet of Things for Smart Cities and Sustainable Building Environment \cite{Alnaser2024AIPowered} &
  2024 &
  \begin{itemize}
    \item Systematic review of 125 papers on AI+IoT‐driven DTs in buildings
    \item Thematic analysis: construction, FM, energy optimization, emerging urban tech
  \end{itemize} &
  Smart Cities; Sustainable buildings; AI; IoT &
  \begin{itemize}
    \item Integration of AI/IoT platforms
    \item Need occupant‐centric, resilient architectures
  \end{itemize}  \\
  \hline

  Comprehensive analysis of digital twins in smart cities: a 4200-paper bibliometric study \cite{El-Agamy2024Comprehensivea} &
  2024 &
  \begin{itemize}
    \item Bibliometric mapping of 4 220 DT papers: trends, authors, clusters
    \item Detailed look at datasets, platforms, performance metrics
  \end{itemize} &
  Smart Cities; Bibliometrics; Digital Twin research &
  \begin{itemize}
    \item Limited deep case‐study insights
    \item Need governance/policy and metric standardization
  \end{itemize}  \\
  \hline

\end{longtable}
\end{small}

\section*{Appendix C: Overview of Data Processing/Fusion Techniques and Simulation Platforms in Selected Digital Twin Studies}
\label{sec:overview-data-processing}

Table~\ref{tab:dt-data-sim-techniques} provides a comparative overview of selected digital twin studies with respect to their data processing or fusion techniques and the simulation platforms they employ. This compilation highlights the diversity of methodological approaches, including traditional statistical models, clustering algorithms, and advanced AI-driven fusion architectures. It also showcases the range of simulation environments used, such as SUMO, CARLA, and custom digital twin frameworks. The table emphasizes emerging trends in real-time analytics, co-simulation, and multi-modal data integration that are shaping the development of intelligent transportation systems.

\begin{small}
\setcounter{table}{0}
\renewcommand{\thetable}{C.\arabic{table}}

\begin{longtable}{|c|p{1.3in}|c|p{2in}|p{1.8in}|}
  \caption{Overview of Data Processing/Fusion Techniques and Simulation Platforms in Selected Digital Twin Studies}
  \label{tab:dt-data-sim-techniques} \\ 
  \hline
  \textbf{No.} 
    & \textbf{Reference} 
    & \textbf{Year} 
    & \textbf{Data Processing/Fusion Techniques} 
    & \textbf{Simulation Platform} 
  \\ \hline
  \endfirsthead
  \multicolumn{5}{c}%
    {{\bfseries \tablename\ \thetable{} -- continued from previous page}} \\
  \hline
  \textbf{No.} 
    & \textbf{Reference} 
    & \textbf{Year} 
    & \textbf{Data Processing/Fusion Techniques} 
    & \textbf{Simulation Platform} 
  \\ \hline
  \endhead
  \hline \multicolumn{5}{r}{{Continued on next page}} \\ 
  \endfoot
  \hline
  \endlastfoot
  1 
    & \cite{Kusic2023digitala} 
    & 2023 
    & Real-time Big Data analytics; continuous fusion of traffic sensor streams; dynamic calibration via SUMO. 
    & SUMO 
  \\ \hline
  2 
    & \cite{Hassani2024systematic} 
    & 2024 
    & Complementary, redundant, and cooperative fusion; multi-level fusion classifications; revised JDL model. 
    & – 
  \\ \hline
  3 
    & \cite{Yilmaz-Niewerth2024Developinga} 
    & 2024 
    & – 
    & Co‐simulation coupling CARLA \& SUMO 
  \\ \hline
  4 
    & \cite{Bamminger2023Evaluationa} 
    & 2023 
    & – 
    & Multi‐body simulation + ML‐based virtual sensor models 
  \\ \hline
  5 
    & \cite{Katsumbe2024Extractiona} 
    & 2024 
    & Word2Vec‐based ML text analytics for unstructured reports and GIS metadata. 
    & – 
  \\ \hline
  6 
      & \cite{Huang2025Generativea} 
      & 2025 
      & Generative spatial AI foundation model (Large Flow Model) for urban digital twins. 
      & Urban Digital Twin framework 
    \\ \hline
    7 
      & \cite{Saroj2023Impacta} 
      & 2023 
      & K‐means clustering; temporal‐neighbor interpolation to impute missing connected‐corridor volume data. 
      & Real‐time connected‐corridor simulation 
    \\ \hline
    8 
      & \cite{Kuraksin2020Integrateda} 
      & 2020 
      & Dynamic Traffic Assignment (DTA) models; ML submodels for traffic performance metrics. 
      & DTALite / NEXTA 
    \\ \hline
    9 
      & \cite{Angelina2024Introducing} 
      & 2024 
      & Fine‐grained energy consumption variables via Python scripts for eco‐driving assessments. 
      & SUMO 
    \\ \hline
    10 
      & \cite{Yadav2025Leasta} 
      & 2025 
      & Least Squares prediction; multi‐criterion asynchronous classifier on acoustic signals. 
      & CupCarbon 
    \\ \hline
    11 
      & \cite{Villanueva-Merino2024Leveraging} 
      & 2024 
      & GIS data integration; AI analytics for age‐friendly urban environment planning. 
      & Local Digital Twin framework 
    \\ \hline
    12 
      & \cite{Papadopoulos2024Lightweighta} 
      & 2024 
      & PCA transformation; DBSCAN clustering for lightweight accident detection from GPS streams. 
      & – 
    \\ \hline
    13 
      & \cite{Helo2024Logistics} 
      & 2024 
      & Real‐time IoT tracking and tracing data integration for logistics visibility. 
      & – 
    \\ \hline
    14 
      & \cite{Louati2025Machinea} 
      & 2025 
      & ARIMA forecasting; ANN predictive modeling for traffic and safety indices. 
      & – 
    \\ \hline
    15 
      & \cite{Lv2022Memoryaugmenteda} 
      & 2022 
      & Memory‐augmented neural networks for dynamic complex image segmentation. 
      & Digital Twin simulation environment 
    \\ \hline
    16 
      & \cite{Aguiar2022MobiWise} 
      & 2022 
      & Multi‐objective optimization; IoT sensor data integration; microservice‐based data‐processing architecture. 
      & Microscopic Traffic Simulator 
    \\ \hline
    17 
      & \cite{Lu2025Modeling} 
      & 2025 
      & STGAT for heterogeneous spatiotemporal pavement‐data fusion. 
      & Digital Twin–enabled highway management 
    \\ \hline
    18 
      & \cite{Varga2023Optimizinga} 
      & 2023 
      & – 
      & Co‐simulation coupling SUMO \& Carla 
    \\ \hline
    19 
      & \cite{Wagner2023SPaTa} 
      & 2023 
      & V2X SPaT/MAP message generation and mixed‐reality virtualization. 
      & Co‐simulation: SUMO, Veins \& OMNeT++ 
    \\ \hline
  20 
    & \cite{Cao2025Optimization} 
    & 2025 
    & Real‐time error coefficient integration with PD control for Pure Pursuit path tracking. 
    & CarSim \& MATLAB/Simulink co‐simulation 
  \\ \hline
\end{longtable}
\end{small}

\section*{Appendix D: Examples of DT applications in smart infrastructure.}
\label{sec:tab-dt-apps}

Table \ref{tab:dt-applications} presents diverse examples of DT applications in smart infrastructure, ranging from traffic management to smart mobility planning.

\setcounter{table}{0}
\renewcommand{\thetable}{D.\arabic{table}}

\begin{table}[h]
\centering
\begin{small}
\caption{Examples of DT Applications in Smart Infrastructure}
\label{tab:dt-applications}
\begin{tabular}{|p{3cm}|p{3.2cm}|p{3.2cm}|p{3cm}|}
\hline
\textbf{Domain} & \textbf{Example Application} & \textbf{Integrated Technologies} & \textbf{Reported Benefits} \\ \hline
Traffic Management & Smart corridor twin for real-time flow and signal control \cite{Safiullin2020Method, Saroj2023Impacta} & IoT sensors; AI (clustering, interpolation); V2X & Optimized congestion management; real-time metrics \\ \hline
Smart Mobility Planning & Citywide mobility twin simulating multimodal transit \cite{Pirdavani2023Application, Stan2024Quantifying} & GIS, transport models; mobile data; AI forecasting & Data-driven planning; proactive congestion mitigation \\ \hline
Road Safety & Intersection safety twin for simulating vehicle-pedestrian interactions \cite{Pugachev2020Features, Ulrich2023Howa} & Cameras, LiDAR; AI-based vision; traffic simulation & Identification of hazards; proactive safety measures \\ \hline
Predictive Maintenance & DT for monitoring infrastructure asset health \cite{Alshuwaikhat2022Analysis, Kerimov2020Model} & Structural sensors; IoT; machine learning & Early fault detection; reduced downtime \\ \hline
Energy Optimization & Integration of BIM and GIS for urban energy management \cite{Jin2024Lightweighting, Huang2025Generativea} & BIM; IoT (smart meters); simulation software & Reduced energy consumption; optimal load balancing \\ \hline
Communication Networks & 5G network digital twin for urban connectivity \cite{Guan2024Key, Wagner2023SPaTa} & 3D GIS; AI-driven ray tracing; network telemetry & Enhanced signal coverage; robust, low-latency connectivity \\ \hline
\end{tabular}
\end{small}
\end{table}

\newpage

\bibliographystyle{unsrtnat}
\bibliography{references} 

\end{document}